\documentclass[twocolumn, pra, showpacs,superscriptaddress]{revtex4}
\usepackage{amssymb}
\usepackage{dcolumn}
\usepackage{bm}
\usepackage{amsmath}
\usepackage{graphicx}
\usepackage{graphics}
\usepackage{float}

\begin{document}

\title{Cavity-enhanced detection of magnetic orders in lattice spin models}
\author{Liping Guo}
\affiliation{Institute of Theoretical Physics, Shanxi University, Taiyuan 030006, P. R.
China}
\author{Shu Chen}
\affiliation{Institute of Physics, Chinese Academy of Sciences, Beijing 100080, P. R.
China}
\author{B. Frigan}
\affiliation{School of Physics, Georgia Institute of Technology, Atlanta, Georgia 30332,
USA}
\author{L. You}
\affiliation{School of Physics, Georgia Institute of Technology, Atlanta, Georgia 30332,
USA}
\author{Yunbo Zhang}
\email{ybzhang@sxu.edu.cn}
\affiliation{Institute of Theoretical Physics, Shanxi University, Taiyuan 030006, P. R.
China}

\begin{abstract}
We develop a general scheme for detecting spin correlations inside a
two-component lattice gas of bosonic atoms, stimulated by the recent
theoretical and experimental advances on analogous systems for a single
component quantum gas. Within a linearized theory for the transmission
spectra of the cavity mode field, different magnetic phases of a
two-component (spin 1/2) lattice bosons become clearly distinguishable. In
the Mott-insulating (MI) state with unit filling for the two-component
lattice bosons, three different phases: antiferromagnetic, ferromagnetic,
and the XY phases are found to be associated with drastically different
cavity photon numbers. Our suggested study can be straightforwardly
implemented with current cold atom experiments.
\end{abstract}

\startpage{1}
\endpage{4}
\maketitle

\section{Introduction}

Atomic quantum gases trapped in optical standing waves have become ideal
systems for implementing lattice spin models after the pioneering
theoretical proposal \cite{Jaksch} and the experimental observation \cite%
{Greiner} of the superfluid (SF) to Mott insulator (MI) transition in the
Bose-Hubbard model. When atoms of two-species or two-components are loaded
into an optical lattice, a variety of more general effective spin models can
be constructed \cite{Duan,Kuklov,leechaohong}, including the well-known
anisotropic Heisenberg XXZ model. The development of noise spectroscopy \cite%
{alt,BN,JN,Mueller,cto,Niu} has provided an astounding breakthrough that
overcomes several significant hurdles in detecting quantum correlations, or
in measuring the second order spin moments for various magnetic phases of
lattice models.

Cold atoms are usually probed with time of flight methods, which measures
the atomic density or matter-wave interference patterns upon being released
from traps and often after significant expansions. The near resonant imaging
light generally destroys the atomic state. Several quantum limited detection
schemes have since been suggested, capable of quantum non-demolition
detections of strongly correlated states in atomic lattice models \cite%
{EckertNP,EckertPRL07}. A very interesting approach relies on the enhanced
detection sensitivity provided by an optical cavity, as was first proposed
by Mekhov \textit{et. al.} \cite{MekhovNP,MekhovPRL07}. The transmission
spectra, calculated to the first order, or within the linear response theory
of the amplitude for the probe field, assumes the initial state of atoms to
remain unchanged when expectation values are taken and carries unambiguous
signatures of magnetic orders in an atomic Bose-Hubbard model.

Several experimental groups have recently succeeded in the difficult first
step of coupling atomic condensates into high Q optical cavities \cite%
{BECQED,gardiner}, highlighting the prospects for creating and detecting
exotic quantum phases of lattice spin models \cite{Larson}. A promising new
direction worthy of theoretical investigation concerns the study of atomic
lattice spin models coupled with optical cavities, generalizing the single
component study \cite{MekhovNP,MekhovPRL07}. Nonlocal quantum spin
correlations of the various magnetic orders could analogously be reflected
through the photon numbers and statistics.

\begin{figure}[tbp]
\centering
\includegraphics[width=3.25in]{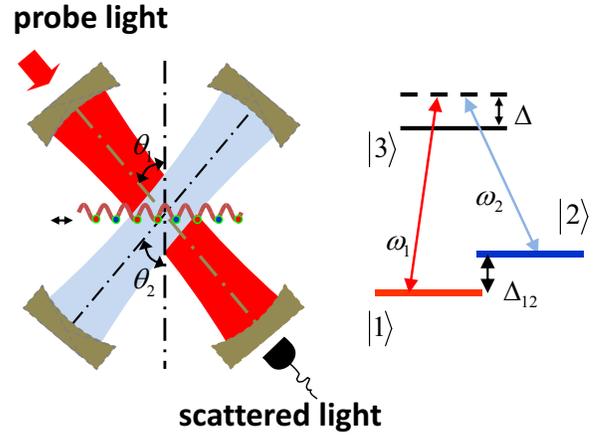}
\caption{(Color online) Schematic illustration of the proposed experimental
setup and the level diagram for a bosonic atom with two states resonantly
coupled to the two cavities.}
\label{fig1}
\end{figure}

This paper describes a scheme for detecting spin correlations in a
two-species or two-component bosonic atom lattice \cite%
{Duan,Kuklov,AltmanNJP,Isacsson}. Our study shows that atomic spin
correlations are faithfully mapped onto the transmission spectra of the
cavity probe field, making them easily diagnosed through cavity QED based
techniques.

\section{Model}

Our model is based on the scattering of two Raman matched incident laser
beams from a lattice of effective spin 1/2 bosonic atoms \cite{JILA,LENS}.
Similar to the original model \cite{MekhovPRL07} for single component
bosons, we consider $N$ atoms with two internal states identically trapped
in an optical lattice with $M$ sites formed by far off-resonant
standing-wave laser beams. As schematically illustrated in Fig. \ref{fig1}, $%
K<M$ lattice sites are located within the overlapped region of the two
fundamental modes of the cavities. We consider two non-degenerate hyperfine
states $\left\vert 1\right\rangle $ and $\left\vert 2\right\rangle $, the
two stable ground states that are coupled to a common excited state $%
\left\vert 3\right\rangle $ with a blue common detuning $\Delta $ and no
differential detuning, forming a Raman coupled $\Lambda$-type atom model.
The resonant cavity modes are denoted by matching labels with frequencies $%
\omega _{l}$ ($l=1,2$). For large detuning $\Delta $, we adiabatically
eliminate the excited state $\left\vert 3\right\rangle $ \cite%
{Alexanian,Gerry} and end up with two-state atoms effectively coupled in the
overlapped region of two optical cavities. For a single atom, the effective
coupling is described by $\Omega a_{1}^{\dagger }a_{2}b_{1}^{\dagger
}b_{2}+h.c.$ and the ac Stark shift becomes $\delta _{l }a_{l}^{\dagger
}a_{l}b_{\sigma=l }^{\dagger }b_{\sigma=l }$ with $\delta
_{l}=g_{l}^{2}/\Delta $ and $\Omega =g_{1}g_{2}/\Delta $. The peak value for
the dipole coupling with their respective cavity mode is denoted by $%
g_{\sigma}$ for transition between $\vert\sigma\rangle \leftrightarrow \vert
3\rangle$. %$E_{\sigma }$ ($\sigma =1,2,3$) denotes the atomic energy for
%the state $\left\vert \sigma\right\rangle$ \cite{Alexanian,Duan}
% and
$b_{\sigma =1,2}$ ($a_{l=1,2}$) denotes the corresponding annihilation
operator for the atom (cavity mode photon).

Following the notations of Ref. \cite{MekhovPRL07}, the Hamiltonian for
effective spin 1/2 bosons in a lattice coupled to two optical cavities takes
the form $H_{B}+H_{I}$, with
\begin{align}
H_{I}& =\sum_{l=1,2}\hbar \omega _{l}a_{l}^{\dagger }a_{l}-i\hbar \eta
\left( a_{1}e^{i\omega _{1p}t}-h.c.\right)  \notag \\
& +\hbar \delta _{1}\sum_{i=1}^{K}\left\vert u_{1}\right\vert
^{2}n_{i1}a_{1}^{\dagger }a_{1}+\hbar \delta _{2}\sum_{i=1}^{K}\left\vert
u_{2}\right\vert ^{2}n_{i2}a_{2}^{\dagger }a_{2}  \notag \\
& +\hbar \Omega \sum_{i=1}^{K}\left( A_{i}a_{1}^{\dagger
}a_{2}b_{i1}^{\dagger }b_{i2}+h.c.\right),  \label{gBH}
\end{align}%
where $n_{i\sigma }=b_{i\sigma }^{\dagger }b_{i\sigma }$ gives the number of
atoms in state $|\sigma \rangle $ at site $i$ and $u_{1,2}(\mathbf{r})$ is
the mode function of the cavity with wave-vector $\mathbf{k}_{1,2}$. The
coefficients $A_{i}(\theta _{1},\theta _{2}) =u_{1}^{\ast} (\mathbf{r}_{i})
u_{2}(\mathbf{r}_{i})$ due to emission/absorption or absorption/emission
cycle are responsible for the geometric dependence of the effective coupling
\cite{MekhovPRL07}.

With atoms assumed to occupy only the lowest Bloch band, our model
generalizes the familiar Bose-Hubbard for two-components: $H_B$ as in Eq.
(1) of Ref. \cite{AltmanNJP} for two species. Following the work of \cite%
{MekhovPRL07}, we perform a linear calculation to the first order in cavity
probe field, thus we leave out the dynamics of how various quantum phases of
the atomic lattice are realized or dynamically created through the tuning of
lattice parameters. This further justifies the neglect of atomic tunneling
as well as the on-site intra- and inter-component interactions. In addition
to the coupling of each atomic component with its corresponding cavity mode,
Raman matched two-photon processes can transfer atoms between the two
effective spin states, unless the atoms are prepared in the so-called dark
state $|\mathrm{dark}\rangle\sim \langle a_2\rangle g_2\left\vert
1\right\rangle -\langle a_1\rangle g_1\left\vert 2\right\rangle$
corresponding to Coherent Population Trapping (CPT) \cite{CPT}. We also
assumed large detuning between cavity and atoms, to keep the actual
excitations low, or negligible; thus any Raman type population transfers
only affect the initial state to higher orders than the linear response
theory calculation we provide. The second term in Eq. (\ref{gBH}) describes
the coherent pumping of cavity 1 at frequency $\omega _{1p}$ with amplitude $%
\eta$.

\section{Semiclassical theory}

We first consider the simplest case with no external pumping on cavity 1,
i.e., $\eta =0$, and assume cavity mode $a_{2}$ to be a classical field, or
a $c$-number amplitude as in Ref. \cite{MekhovPRL07}. In the frame rotating
with frequency $\omega _{2}$, $a_{1}$ evolves in time according to the
Heisenberg equation%
\begin{equation}
\dot{a}_{1}=-i(\Delta _{12}+\delta
_{1}\sum_{i}^{K}|u_{1}|^{2}n_{i1})a_{1}-i\Omega \hat{D}a_{2}-\kappa a_{1},
\end{equation}%
where $\Delta _{12}=\omega _{1}-\omega _{2}$ and $\kappa $ denotes the
cavity decay rate and is put in by hand. Its corresponding Langevin noise is
neglected. We have defined the analogous operator $\hat{D}%
=\sum_{i=1}^{K}A_{i}S_{i}^{-}$, in terms of the effective lattice spin
operators $S_{i}^{-}=b_{i1}^{\dagger }b_{i2}$ and $S_{i}^{+}=\left(
S_{i}^{-}\right) ^{\dag }$, which obey the standard commutation relation at
the same site and commute with each other on different sites. Neglecting the
presumably much smaller cavity field induced ac Stark shift in comparison to
$\Delta _{12}$ or $\kappa $ \cite{MekhovNP}, $a_{1}$ and the photon number
is easily obtained as
\begin{equation}
a_{1}=C\hat{D},\qquad a_{1}^{\dagger }a_{1}=\left\vert C\right\vert ^{2}\hat{%
D}^{\dag }\hat{D},
\end{equation}%
where $C=-i\Omega a_{2}/(i\Delta _{12}+\kappa )$. The photon number $\langle
a_{1}^{\dagger }a_{1}\rangle $, clearly provides information about the spin
correlation in the two-component bose lattice through the moments associated
with the same site $\langle S_{i}^{+}S_{i}^{-}\rangle $ and between the
different sites $\langle S_{i}^{+}S_{j}^{-}\rangle $. The angular dependence
can become totally different due to the geometric coefficients $A_{i}(\theta
_{1},\theta _{2})$. Within the linear response, the above averages are
expectation values with respect to whatever initially prescribed atomic
ground state.

\begin{table}[tbp]
\centering
\begin{tabular}{|c|c|c|}
\hline
& $\langle a_{1}^{\dagger }a_{1}\rangle _{\theta _{1}=0}$ & $\langle
a_{1}^{\dagger }a_{1}\rangle _{\theta _{1}=\pi /2}$\hfill \\ \hline
$\text{AF}$ & $K|C|^{2}/2$ & $K|C|^{2}/2$ \\ \hline
$\text{FM}$ & $0$ & $0$ \\ \hline
$\text{XY}$ & $(K+3K^{2}) |C|^{2}/16$ & $K|C|^{2}/16$ \\ \hline
$\text{SF}$ & $n_{2}(n_{1} K+1) K|C|^{2}$ & $n_{2}K|C|^{2}$ \\ \hline
\end{tabular}%
\caption{Cavity 1 photon number for the four quantum phases of the
two-component Bose-Hubbard model at the diffraction maxima (minima) with $%
\protect\theta_1=0$ ($\protect\theta_1=\protect\pi/2$) and $\protect\theta%
_{2}=0$. For the XY phase $\protect\theta_A=\protect\theta_B=\protect\pi/3$.}
\label{table1}
\end{table}

The quantum phases for a two-component lattice bosons at commensurate
fillings have attracted significant attention \cite{AltmanNJP,Isacsson}. The
phase diagram consists of (1), 2MI where both boson components are in the MI
phase; (2), SF+MI where one is SF and the other is MI; and (3), 2SF where
both components are SF. Deep inside the MI phase the ground state of the
system may be characterized by filling the lattice site with even or odd
numbers of atoms \cite{Isacsson}. In addition to the usual even filling
phase with $n_{1}=n_{2}$, a particularly interesting phase arises when the
total filling factor is odd, especially at unit filling, i.e., for $%
n_{1}+n_{2}=1$. This exotic phase has been extensively studied \cite%
{Duan,Kuklov,Isacsson,AltmanNJP} by adopting a trial wave function $%
\vert\Psi_{MI}\rangle =\prod_{i\in A,j\in B} \vert\psi_{A}\rangle _{i}\vert
\psi_{B}\rangle_{j}$, which is of a form composed of two sublattices $A$ and
$B$ with $\vert\psi _{A,B}\rangle =\cos ({\theta_{A,B}}/{2})\vert 1,0\rangle
+e^{i\phi _{A,B}}\sin ({\theta _{A,B}}/{2})\vert 0,1\rangle $. $\vert
n_{1},n_{2}\rangle_{i}$ denotes the state with $n_{1}$ $(n_{2})$ number of
component-1 (-2) atoms at site $i$ and $\theta$s and $\phi$s are variational
parameters. Three types of spin exchange interactions are identified: (I),
anti-ferromagnetic phase (AF) with $\theta_{A}=0 (\pi)$ and $\theta_{B}=\pi
(0)$; (II), ferromagnetic phase (FM) with $\theta_{A}=\theta_{B}=0$; and
(III), XY phase with $\theta_{A}=\theta_{B}\neq 0$. The 2SF phase, whose
quantum state is $\Psi _{SF}\sim (\sum_{i}b_{i1}^{\dagger
})^{N_{1}}(\sum_{j}b_{j2}^{\dagger })^{N_{2}}\vert 0\rangle $ with $N_{1,2}$
the total number of component-1 (-2) atoms \cite{Rodriguez}, will serve as a
reference for presenting our results.

The scattered photons are explicitly tabulated in Table \ref{table1}. For a
1D optical lattice of a spatial period $d=\lambda /2$ and with atoms trapped
at sites centered at $x_{j}=jd$, the mode functions are $u_{1,2} (\mathbf{r}%
_{j}) =\exp(ij\vert \mathbf{k}_{1,2}\vert d\sin \theta _{1,2})$ for a
traveling wave and/or $u_{1,2}\left( \mathbf{r}_{j}\right) =\cos(ij\vert
\mathbf{k}_{1,2}\vert d\sin \theta _{1,2})$ for a standing wave form. Atoms
in the FM phase do not scatter because the two coupling paths to the excited
state $\left\vert 3\right\rangle$ destructively cancels as in the dark
state. For the notation we use, the FM state corresponds to all atoms
staying in state $\left\vert 1\right\rangle$, then a semi-classical light
amplitude $\left<a_2\right>g_2$ clearly will not be able to cause any
scattering. While the initial atomic states of the AF and XY phases under
the single excitation of a semi-classical light are not any more dark
states, they will scatter. These features thus completely characterize the
many-body spin correlations of the quantum phases for the two-component
Bose-Hubbard model.

\begin{figure}[tbp]
\includegraphics[width=3.4in]{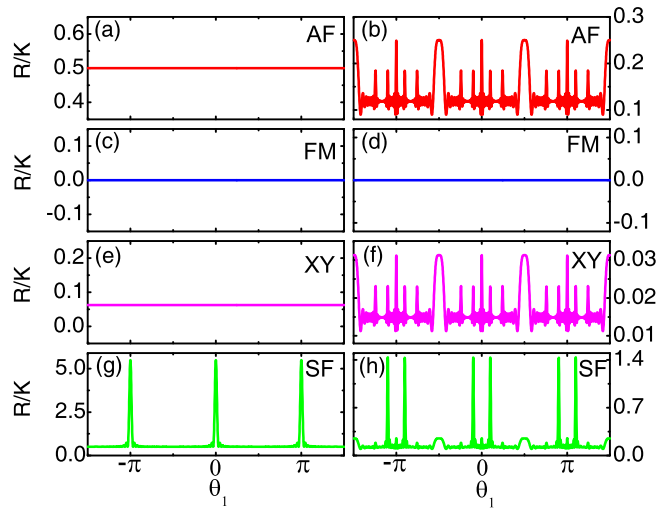}
\caption{(Color online) The angular distribution of $R(\protect\theta _{1},%
\protect\theta_2)$ for the four quantum phases evaluated for different
choices of cavity mode functions: the left (right) panels are for two
traveling (standing) waves and for $\protect\theta_2=0$ ($\protect\theta%
_2=0.1 \protect\pi$). We have assumed $N=M=2K=40$ and in the SF phase $%
n_{1}=n_{2}=1/2$. For the XY phase $\protect\theta_A=\protect\theta_B=%
\protect\pi/3$.}
\label{fig2}
\end{figure}

To map quantum fluctuations of lattice spins faithfully onto the probe
cavity photon statistics, we define a noise function $R(\theta_{1},%
\theta_{2})=\langle D^{\dag }D\rangle -\langle D^{\dag }\rangle\!\langle
D\rangle $, whose angular distribution is compared in Fig. \ref{fig2} for
all four quantum phases. The structure in the angular distribution comes
from the summation of the geometric coefficients from different sites,
reflecting both the on-site and off-site lattice spin correlations. In the
SF phase with $n_{1}=n_{2}=1/2$, the respective noise functions are
completely different for the two choices of cavity modes. For the traveling
wave, the noise function is zero for the FM phase, but takes nonzero values
and is isotropic for the XY and the AF phases. The angular dependence for
the standing wave mode case is richer than that for the traveling waves. The
structures in the angle dependence can be attributed to dependence on the
summation of the geometric coefficients, and physically due to both on-site
and off-site lattice spin correlations.

\section{Quantized model}

We next consider the more general case with coherent pumping for cavity 1 at
frequency $\omega _{1p}$ \cite{MekhovPRL07}. The dissipations for both
cavities are assumed the same with the associated Langevin noise terms
neglected in the Heisenberg operator equations. Within a linearized
calculation, we decorrelate the atomic and field operators and replace in
the Heisenberg equations for $a_{1,2}$ the atomic operators by their
respective expectation values, which leads to $\langle a_{l}^{\dag }\rangle
\langle a_{l}\rangle =|\langle a_{l}\rangle |^{2}$. To simplify our result,
we further assume $|u_{1,2}(\mathbf{r}_{i})|^{2}=1$, which occurs for the
diffraction maxima with $A_{i}=1$ at $\theta _{1}=0$ or the minima with $%
A_{i}=(-1)^{i}$ at $\theta _{1}=\pi /2$ when the 1D lattice is lined up at $%
\theta _{2}=0$. The cavity photons are found to be
\begin{equation}
\langle a_{1}^{\dag }\rangle \langle a_{1}\rangle =\eta ^{2}(\kappa
^{2}+\zeta _{2}^{2})/B,\quad \langle a_{2}^{\dag }\rangle \langle
a_{2}\rangle =\eta ^{2}\alpha ^{\ast }\alpha /B,  \label{aa}
\end{equation}%
where $B=\kappa ^{4}+\kappa ^{2}(\zeta _{1}^{2}+\zeta _{2}^{2}+2\alpha
^{\ast }\alpha )+(\zeta _{1}\zeta _{2}-\alpha ^{\ast }\alpha )^{2}$, $\alpha
=\Omega \sum_{i}^{K}A_{i}\langle S_{i}^{-}\rangle $, and $\zeta _{l}=\Delta
_{lp}+\delta _{l}\sum_{i}^{K}\langle n_{il}\rangle $. The detuning $\Delta
_{lp}=\omega _{l}-\omega _{1p}$ are assumed the same for $l=1,2$ because $%
\Delta _{12}\ll \omega _{1,2}$. If the cavity coupling is assumed identical,
we end up with $\delta _{1,2}=\Omega =\delta $. Equation (\ref{aa}) shows
that probe photon numbers depend on the average values of on-site atom
numbers $\langle n_{i\sigma }\rangle $ and the lattice spin operators $%
\langle S_{i}^{\pm }\rangle $. A crucial term for spin correlation $\alpha
^{\ast }\alpha $ appears in the expression for $\langle a_{2}^{\dag }\rangle
\langle a_{2}\rangle $. At the diffraction minima or maxima $\alpha =0$ so
that no photon will be detected from cavity 2 except for the XY phase. This
then allows for simplified expressions of the scattered photon numbers $%
\langle a_{1}^{\dag }\rangle \langle a_{1}\rangle $ from the AF and FM
phases into $\langle a_{1}^{\dag }\rangle \langle a_{1}\rangle =\eta
^{2}/(\kappa ^{2}+\zeta _{1}^{2})$, which only depends on the detuning $%
\Delta _{1p}$ and atom numbers for component-1 in the overlapped $K$-sites $%
N_{1}^{K}=\sum_{i}^{K}\left\langle n_{i1}\right\rangle $. When $\alpha \neq 0
$, however, $\langle a_{1}^{\dag }\rangle \langle a_{1}\rangle $ for the XY
phase at the diffraction maxima depends on two parameters $\zeta _{1}$ and $%
\zeta _{2}$ including the detunings $\Delta _{1p}$, $\Delta _{2p}$, and the
number of atoms for both components in the overlapped region of $K$-sites.
Measuring photon numbers $\langle a_{1}^{\dag }\rangle \langle a_{1}\rangle $
thus gives sufficient information to distinguish magnetic orders or quantum
phases of the two-component Bose-Hubbard model.

An especially interesting property concerns the dependence of the probe
photon numbers on the detuning $\Delta_{1p}$, as is illustrated in Fig. \ref%
{fig3} for the four quantum phases. For the FM and AF phases, we find
Lorentzians with width $\kappa $ and shifted by $\delta N_{1}^{K}$ as in the
classical result of a single component Bose-Hubbard model \cite{MekhovPRL07}%
. In contrast, for the SF phase the photon number distribution is an
envelope of a comb for a good cavity ($\kappa =0.1\delta ) $ while a smooth
broadened contour for a bad cavity ($\kappa =\delta )$. In the SF case,
individual atoms are completely delocalized over all sites causing
significant number fluctuations over each site within the $K$-site region.
The corresponding quantum state is a superposition of Fock states containing
all possible distributions of $N_{1}^{K}$ atoms for component-1 at $K $
sites, which gives rise to scattering terms from all possible atomic
distributions. For the XY phase, the double peaked feature provides evidence
for different population of atoms in the two internal states, with the
relative heights of the two peaks being controlled by the variational
parameters $\theta_{A,B}$. This structure in the easy plane XY phase is
essentially identified with the so-called superfluid counterflow (SCF)
phase, which can be qualitatively understood as a paired superfluid vacuum
(PSF) phase, a strongly correlated superfluid ground state already predicted
from numerical simulations \cite{Kuklov}. These distinct features of the
transmission spectrum we discuss for the various quantum phases form the
basis for easily detecting and differentiating the corresponding magnetic
orders in the two-component Bose-Hubbard model.

\begin{figure}[tbp]
\includegraphics[width=3.35in]{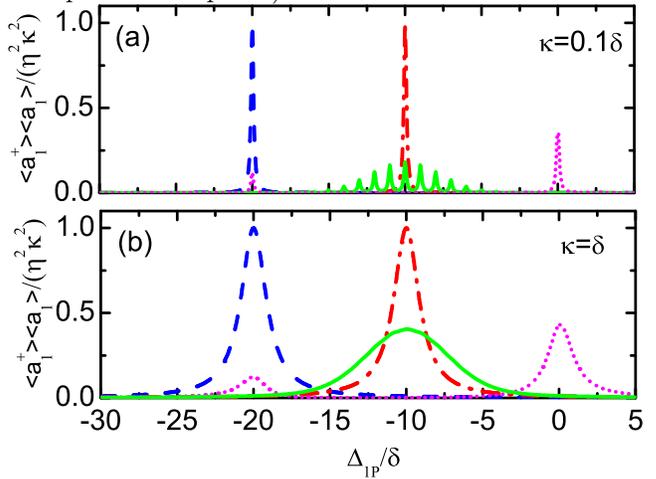}
\caption{(Color online) Cavity 1 photon numbers as a function of
cavity-probe detuning for the four quantum phases: AF (red dashed dot), FM
(blue dashed), XY (pink dotted), and SF (green solid). In our simulation we
use $K=20$ for all phases and in the SF phase $n_{1}=n_{2}=1/2$. For the XY
phase $\protect\theta_A=\protect\theta_B=0.6\protect\pi$.}
\label{fig3}
\end{figure}

Like the original cavity scheme of Mekhov et. al. \cite{MekhovNP,MekhovPRL07}%
, the scheme we propose, is constructed to detect high order moments. The
different phases (in the sense of quantum states of matter) of a
two-component lattice bose gas are resolved from the statistics of scattered
photons or pseudo-spins. In this sense, it is analogous to the so-called
noise spectroscopy of quantum gases \cite{alt,Mekhov2}, albeit somewhat
superior due to the enhanced collection efficiency aided by cavities. The
Ramsey spectroscopy \cite{Ramsey}, as proposed by Kuklov, measures the first
order moments of atomic pseudo-spins. The SCF state or the paired
condensation phase is a special case, where the order parameters are simply
field operators themselves. Thus their presence can be probed by the Ramsey
spectroscopy measurement of the relative phase (in the sense of amplitude
and phase).

\section{Conclusions}

In summary we have generalized the model of a single component atomic
lattice gas described by the Bose-Hubbard model coupled to near resonant
optical cavities to the case of a two-component Bose-Hubbard model. We have
shown conclusively through the probe cavity photon numbers and its spectra
dependence on various system parameters that different quantum phases of the
two-component Bose-Hubbard model can be easily distinguished and confirmed.
Our results shine new light on atomic lattice gases coupled to cavity QED
systems.

\begin{acknowledgments}
This work is supported by NSF of China under Grant No. 10774095,
10434080 and 10574150, and the 973 Program under Grant No.
2006CB921102. L.Y. acknowledges support from ARO.
\end{acknowledgments}

\end{document}